\documentclass[conference]{IEEEtran}
\IEEEoverridecommandlockouts

\usepackage{amsmath,amsfonts}
\usepackage{algorithmic}
\usepackage{algorithm}
\usepackage{array}
\usepackage{textcomp}
\usepackage{stfloats}
\usepackage{url}
\usepackage{verbatim}
\usepackage{graphicx}
\usepackage{cite}
\usepackage{graphicx} 
\usepackage{acronym}
\usepackage{amsmath}
\usepackage{amsfonts}
\usepackage{xcolor}
\usepackage{fancyhdr}
\usepackage{url}
\usepackage{tabularray}
\usepackage{multirow}
\usepackage{booktabs}
\usepackage{amssymb}
\usepackage{url}
\usepackage[caption=false,font=footnotesize]{subfig}

\acrodef{IIoT}[IIoT]{Industrial Internet of Things}
\acrodef{CIR}[CIR]{channel impulse response}
\acrodef{DCIR}[DCIR]{Directional Channel Impulse Response}
\acrodef{RSRP}[RSRP]{Reference Signal Received Power}
\acrodef{MPC}[MPC]{multipath component}
\acrodef{DL}[DL]{Deep Learning}
\acrodef{QoS}[QoS]{Quality of Service}
\acrodef{PQoS}[PQoS]{Predictive Quality of Service}
\acrodef{DPQoS}[DPQoS]{Distributed Predictive Quality of Service}
\acrodef{LOS}[LOS]{line-of-sight}
\acrodef{NLOS}[NLOS]{non-line-of-sight}
\acrodef{ODE}[ODE]{Ordinary Differential Equation}
\acrodef{GLNN}[GLNN]{Graph-Liquid Neural Network}
\acrodef{RIS}[RIS]{Reflective Intelligent Surface}
\acrodef{QoS}[QoS]{Quality of Service}
\acrodef{PQoS}[PQoS]{Predictive Quality of Service}
\acrodef{DPQoS}[DPQoS]{Distributed Predictive Quality of Service}
\acrodef{ML}[ML]{Machine Learning}
\acrodef{GML}[GML]{Machine Learning on Graphs}
\acrodef{UE}[UE]{User Equipment}
\acrodef{BS}[BS]{Base Station}
\acrodef{NN}[NN]{Neural Network}
\acrodef{RNN}[RNN]{Recurrent Neural Network}
\acrodef{CNN}[CNN]{Convolutional Neural Network}
\acrodef{LTC}[LTC]{Liquid Time Constant Network}
\acrodef{MIMO}[MIMO]{Multiple-Input-Multiple-Output}
\acrodef{GNN}[GNN]{Graph Neural Network}
\acrodef{STGNN}[STGNN]{Spatio-Temporal Graph Neural Network}
\acrodef{PPP}[PPP]{Poisson Point Process}
\acrodef{LSTM}[LSTM]{Long-Short Term Memory}
\acrodef{PL}[PL]{path-loss}
\acrodef{CDF}[CDF]{cumulative distribution function}
\acrodef{RMS}[RMS]{root mean square}
\acrodef{SC}[SC]{Sparse clutter}
\acrodef{DC}[DC]{Dense clutter}
\acrodef{RX}[RX]{receiver}
\acrodef{TX}[TX]{transmitter}
\acrodef{PDP}[PDP]{power-delay profile}
\acrodef{SNR}[SNR]{signal to noise ratio}
\acrodef{CSI}[CSI]{channel state information}
\acrodef{sub-THz}[sub-THz]{sub-terahertz}
\acrodef{JSAC}[ISAC]{Integrated Sensing and Communication}
\acrodef{BW}[BW]{bandwidth}
\acrodef{FMCW}[FMCW]{frequency modulated continous wave}
\acrodef{FSPL}[FSPL]{free-space path loss}
\acrodef{RSSI}[RSSI]{received signal strength indicator}

\def\BibTeX{{\rm B\kern-.05em{\sc i\kern-.025em b}\kern-.08em
    T\kern-.1667em\lower.7ex\hbox{E}\kern-.125emX}}

\fancypagestyle{firstpage}
{
    \fancyhf{}
    \fancyhead[C]{The paper has been presented at the 2025 IEEE Future Networks World Forum (FNWF),\\ Bangalore, India, November 10-12, 2025, \url{https://doi.org/10.1109/FNWF66845.2025.11317573}}
    \fancyfoot[C]{This research was funded in part by the National Science Center (NCN), Poland, grant no. 2021/43/I/ST7/03294 (MubaMilWave). For this purpose of Open Access, the author has applied a CC-BY public copyright license to any Author Accepted Manuscript (AAM) version arising from this submission.}
}

\begin{document}

\title{Angularly-Resolved 3D Foliage Modeling and Measurements at 60 and 80\,GHz: From Stochastic Geometry to Deterministic Channel Characterization\\

}

\author{
    \IEEEauthorblockN{
    Jiri Blumenstein\IEEEauthorrefmark{1}, 
    Radek  Zavorka\IEEEauthorrefmark{1}, Josef Vychodil\IEEEauthorrefmark{1}, Tomas Mikulasek\IEEEauthorrefmark{1}, Jarosław Wojtuń\IEEEauthorrefmark{3}, Jan M. Kelner\IEEEauthorrefmark{3},\\ Cezary Ziółkowski\IEEEauthorrefmark{3}, Rajeev Shukla\IEEEauthorrefmark{2}, Markus Hofer\IEEEauthorrefmark{4}, Thomas Zemen\IEEEauthorrefmark{4}, Christoph F Mecklenbrauker\IEEEauthorrefmark{6},\\ Aniruddha Chandra\IEEEauthorrefmark{2}, Ales Prokes\IEEEauthorrefmark{1}} 
    
    \IEEEauthorblockA{
        \IEEEauthorrefmark{3}Institute of Communications Systems, Military University of Technology, Warsaw, \textit{Poland};
        \IEEEauthorrefmark{2}ECE Department,\\ National Institute of Technology, Durgapur, \textit{India};
        \IEEEauthorrefmark{4}AIT Austrian Institute 
        of Technology, Vienna, \textit{Austria};\\
        \IEEEauthorrefmark{6}Institute of Telecommunications, TU Wien, Vienna, \textit{Austria};
        \IEEEauthorrefmark{1}Brno University of Technology, Brno, \textit{Czechia}
    }
}

\maketitle
\thispagestyle{firstpage}

\begin{abstract}
In this paper, we show a stochastic approach to generate a 3D model of a foliage, which is then used for deterministic ray-tracing channel modeling. This approach is verified by a measurement campaign at 60 and 80\,GHz with 2\,GHz bandwidth. The wireless channel is characterized by path-loss and RMS delay spread and we show the angular dependency of those parameters when the receiver is placed on a half-circle around the tree. Besides electromagnetic material properties, the 3D model is characterized by several interpretable parameters, including tree volume, leaf size, leaf density, and the tree crown shape parameter. 

\end{abstract}

\begin{IEEEkeywords}
ray-tracing, channel modeling, channel sounding, millimeter waves.
\end{IEEEkeywords}

\section{Introduction}
\begin{figure}[!t]
\centering
\includegraphics[width=3in]{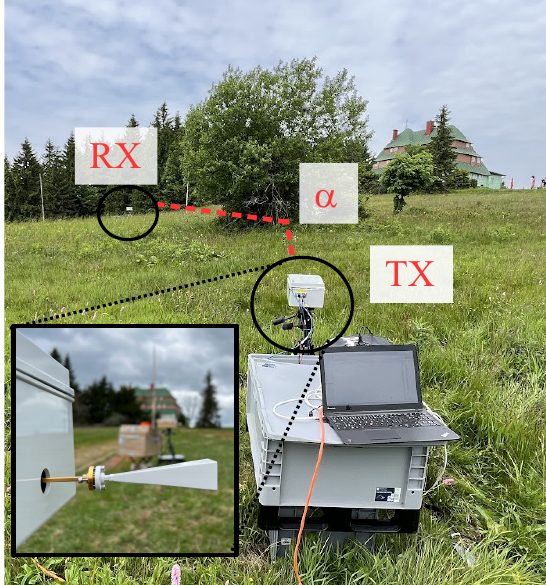}
\caption{{Photo of the 60 and 80\,GHz channel sounder. The \ac{RX} is located on the left side from the measured foliage. The distance from the TX to the tree and from the tree to the RX  is 15\,m.}}
\label{fig:foto}
\end{figure}

\begin{figure}[!t]
\centering 
\includegraphics[width=3in]{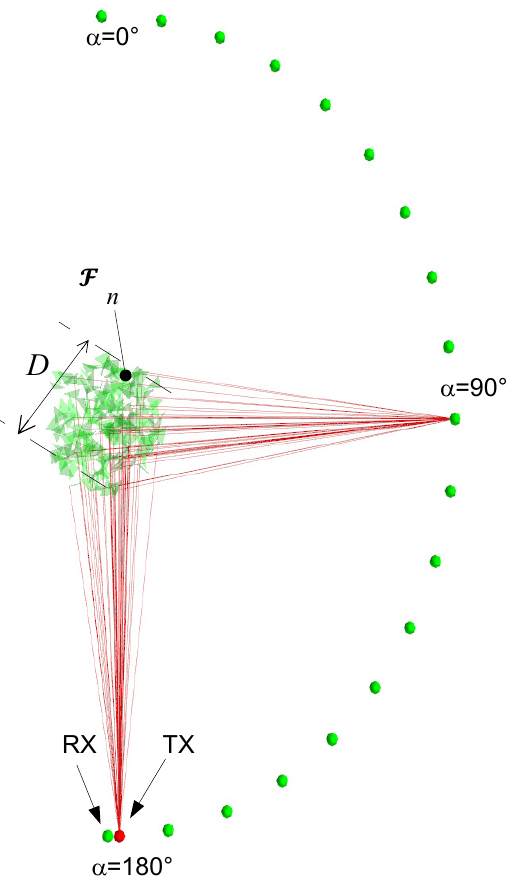}
\caption{{ 
Top-down view on the ray-tracing scenario depicting the half-circular placement of receivers. For clarity, we depict rays only for $\alpha=90$°.  $D$ stands for three crown diameter and the triangular element $\mathcal{F}_n$ represents $n$-th face. Please note that the \ac{LOS} is suppressed and not visualized. We assume isotropic radiation. The transmitter is slightingly moved sideways to allow for visualization.
}}
\label{fig:RTscene}
\end{figure}

The millimeter-wave band, spanning the electromagnetic spectrum from 30 to 300\,GHz, offers extraordinary bandwidth and, consequently, high potential data rates. It is also well known that path loss is a function of frequency. Compared to lower frequencies, millimeter-wave channels thus exhibit a higher path loss. However, this can be compensated for by utilizing higher-order antenna arrays with the same physical area as antennas operating at lower frequencies \cite{rappaport}. The more problematic part of the millimeter-wave band is a low diffraction contribution and high penetration losses \cite{8030501}. Thus, in this paper we concentrate on the problematic situations when the \ac{LOS} is obstructed by  vegetation and only the reflection and/or the diffraction contributions are available (please see Fig. \ref{fig:foto} and \ref{fig:RTscene}).

Either derived heuristically, from measurement datasets or from geometrical representations, various channel modeling techniques have been explored in the literature \cite{8207426}, but a complete review of all the modeling methods exceeds the scope of this paper. Instead, we concentrate on deterministic methods, particularly ray-tracing-based site-specific channel modeling, which offers interpretable and visualizable~\ac{CSI} \cite{zemen_1}.

Let us note that the accuracy of the derived \ac{CSI} heavily relies on the precision of the 3D environment models used \cite{10133177}. This reliance is especially critical at millimeter-wave frequencies, where the wavelength is in the millimeter range, rendering even minor environmental details vital for model accuracy. 

A method for 3D foliage modeling presented in \cite{9411133} utilizes hollow geometric shapes to represent trees and foliage. This approach applies extra attenuation depending on the distance rays traverse through foliage blocks. 

Similarly, in \cite{s21124112}, trees are modeled as hollow cubic forms, with the outer foliage shape (the envelope) significantly simplified. For wide-band signals with high spatial resolution, the lack of ray interactions within the foliage volume may lead to unwanted artifacts in the \acp{CIR}. Thus, in \cite{9262930}, the cubic volume is divided into voxels, whereas in \cite{9524481}, trees are represented by canopy and trunk, each modeled as concentric cylinders. In \cite{9569961}, vegetation scattering is represented through a hybrid approach combining physics-based modeling with data-driven elements. The foliage is described as a dielectric slab populated with randomly oriented leaves (modeled as disks) and branches (modeled as cylinders), which contribute to both scattering and attenuation of the electromagnetic field. In \cite{8668814}, scattering within vegetation is instead approximated by placing point scatterers inside a tree volume, typically defined by simple geometric shapes such as spheres or cylinders.

In \cite{zemen_2}, the authors employ point scatterers to model roadside scattering elements. These scatterers are treated as infinitesimal points, lacking physical thickness or randomized orientation—unlike actual leaves in a natural tree. This abstraction limits the physical realism of the model and affects both the visualization and interpretability~of~ray-tracing~results.

In this paper, we present a methodology for the generation of the 3D foliage model and subsequent ray-tracing simulation. The fundamentals of this approach are described in \cite{blumen_awpl} and  in \cite{10703038} we provide in-depth details of the measurement campaigns including the description of the channel sounders. The contribution of this paper is as follows:
\begin{itemize}
    \item Results of angularly-resolved simulations of the path-loss and \ac{RMS} delay spread of a wireless propagation through a foliage.
    \item Comparison and validation of the simulated and measured channels at 60 and 80\,GHz with 2\,GHz bandwidth.
\end{itemize}

\section{Stochastic 3D Foliage Modeling methodology}
The 3D model consists of vertices, where each triplet of vertices forms a triangular face. A set of these triangular faces represents scattering objects that are randomly positioned and oriented within a predefined area - we can view this as a tree crown envelope. Such a cloud of randomly placed and rotated scattering faces serves as an input for the ray-tracing simulation. In the following section, we provide a more rigorous description of the 3D model generation process.

\subsection{Creation of the 3D foliage model}
The basis for modeling the shape (envelope) of the tree crown is a pair of sets, the set of vertices $\mathcal{V'}$ and the set of faces~$\mathcal{F'}$. We write:
\begin{equation}
\Gamma' = (\mathcal{V'}, \mathcal{F'}), \quad \text{where } \mathcal{V'} \subset \mathbb{R}^3.
\end{equation}
Specifically, the set of vertex positions is written as:
\begin{equation}
\mathcal{V'} = \{ \mathbf{v'}_1, \mathbf{v'}_2, \ldots, \mathbf{v'}_{3M} \}, \quad \mathbf{v'}_i \in \mathbb{R}^3    
\end{equation}
and the set of triangular surfaces is given by:
\begin{equation}
\mathcal{F'} = \{ \mathcal{F}'_1, \mathcal{F}'_2, \ldots, \mathcal{F}'_M \},
\end{equation}
where $M$ is the number of triangles.

For simplicity, we assume that each vertex lies  on the unit sphere, $\|\mathbf{v}_i\| \approx 1 $. To obtain the irregular shape of the tree crown, we utilize a point-wise deformation function $P: \mathcal{V}' \rightarrow \mathcal{V}$, which deforms the initial shape $\Gamma'$ by adding a random displacement $\boldsymbol{\delta}_i $ sampled from a 3D normal distribution. The vertices are then given~as:
\begin{equation}
{\mathbf{v}}_i = \mathbf{v'}_i + \boldsymbol{\delta}_i, \quad \boldsymbol{\delta}_i \sim \mathcal{N}(\mathbf{0}, \xi^2 \mathbf{I}),
\end{equation}
where $\xi$ is a user-defined parameter controlling the deformation strength.

In order to set the volume of the tree crown, we determine the ratio of the desired volume $V_\mathrm{D}$ and the initial volume $V_\mathrm{I}$ as: $\sigma = \left( \frac{V_{\text{D}}}{V_I} \right)^{1/3}$. Then, the vertices are scaled according: $\mathbf{v}_i^{\text{D}} = \sigma \cdot {\mathbf{v}}_i$. With this scaling operation we ensure that the resulting shape $\Gamma=(\{\mathbf{v}_1^{\text{final}},\mathbf{v}_2^{\text{final}},\ldots,\mathbf{v}_N^{\text{final}} \},\mathcal{F'})$ approximately matches the desired volume of the tree crown.

At this stage, the tree crown is ready to be filled with scattering triangular faces $\mathcal{F}\subset \mathbb{N}^3$, creating an  stochastic internal structure of the foliage.  Each $\mathcal{F}$ is defined in a local coordinate system centered at the origin and its vertices are written as:
\begin{equation}
\mathbf{p}_1\!=\!\left(0, -\frac{l}{\sqrt{3}}, 0\right), \mathbf{p}_2\!=\!\left(\frac{l}{2}, \frac{l}{2\sqrt{3}},\!0\right), \mathbf{p}_3\!=\!\left(-\frac{l}{2}, \frac{l}{2\sqrt{3}}, 0\right)
\end{equation}
where $l = \sqrt{ \frac{4A}{\sqrt{3}} }$ is  the side length of the equilateral triangle with user-defined area $A$, representing the actual scattering face.
The next step is that each face is randomly rotated utilizing a Rodrigues' rotation  matrix $\mathbf{R} \in \mathbb{R}^3$.

The triangles $\mathcal{F}$ are defined by local coordinates $\mathbf{p}_j \in \mathbb{R}^3$ (for $j = 1,2,3$), and their rotated counterparts are defined as: 
\begin{equation}
\mathbf{p}_j' = \mathbf{R} \cdot \mathbf{p}_j.    
\end{equation}
 In the next step, the rotated triangular face is translated to a random position $\mathbf{s}_j$ utilizing a random offset $\mathbf{c}~\in~\mathbb{R}^3~\text{ such that } \mathbf{c}~\sim~\mathcal{U}(\Gamma)
\quad \text{and} \quad \mathbf{c} \in \Gamma$.  Here, $\mathcal{U}(\Gamma)$  denotes a uniform distribution  within $\Gamma$, meaning that every point within the volume has the same probability of occurrence. Now, the rotated and translated vertices are given~as:
\begin{equation}
\mathbf{s}_{i1} = \mathbf{c}_i + \mathbf{R} \cdot \mathbf{p}_1, \quad
\mathbf{s}_{i2} = \mathbf{c}_i + \mathbf{R} \cdot \mathbf{p}_2, \quad
\mathbf{s}_{i3} = \mathbf{c}_i + \mathbf{R} \cdot \mathbf{p}_3
\end{equation}
This process of filling the envelope $\Gamma$ could be repeated in order to satisfy the density of the faces $\rho$. 

\section{Ray-tracing and measurements}
In this section we describe both ways of gathering  data from the measurement campaign and from the subsequent ray-tracing simulations.

\subsection{Ray-tracing}
The ray-tracing simulation of wireless propagation through the 3D foliage model is performed using the open-source ray tracer Sionna, version 1.0.2 \cite{sionna}. The material parameter settings are based on the ITU-R P.833 recommendation~\cite{ITU-RP833}. We use a scattering coefficient $\mu_s = 0.50$, relative permittivity of $\epsilon_r = 17$ and conductivity $\kappa = 0.05$\,S/m. Utilizing the pulse-shaping, we reduce the \ac{BW} to 2\,GHz. Table \ref{tab:parameters} lists the 3D model parameters. 

As for the computational cost, the model consisted of hundreds of triangles/faces which is the main factor determining the computational duration. In the presented cases, the simulation took milliseconds on a regular office laptop with a Ryzen 5 processor and without a dedicated GPU. 

\renewcommand{\arraystretch}{1.3}  
\begin{table}[h!]
\centering
\begin{tabular}{|c|l|l|}
\hline
\textbf{Parameter} & \textbf{Description}\\
\hline
$A$ [$\mathrm{m}^2$] & Area of each internal triangle \\
\hline
$V_{\text{D}}$ $[\mathrm{m}^3]$ & Target volume of the outer shape \\
\hline
$\xi [-]$ & Perturbation strength (standard deviation of noise)\\
\hline  
$\rho$ [$\mathrm{triangles/m}^3$] & Internal triangle density \\
\hline
$r_{\text{seed}}$ $[-]$ & Random seed (optional, for reproducibility)\\
\hline
\end{tabular}
\caption{{Table of user-defined parameters for the tree-crown shape generation and internal triangle placement.}}
\label{tab:parameters}
\end{table}

\subsection{Measurement campaign}
The measurement environment is depicted in Fig. \ref{fig:foto}, where we see the tree under test. It is a deciduous tree with rather small leaves  (4-5\,cm of length) that is measured during the summer. The \ac{TX} is located 15\,m in front of the tree and its horn antenna with 24.8\,dBi gain is always pointed towards the tree and parallel to the ground. The same holds for the \ac{RX}, which was moved around the tree on a half-circular trajectory, while its horn antenna with 24.8\,dBi gain  is also parallel to the ground and pointed towards~the~tree. 

In \cite{10703038}, the outline of the \ac{RX} and \ac{TX} measurement positions is provided. For the verification of the ray-tracing outcome, we utilize the positions RX2 at $\alpha=180$° and RX4 at $\alpha=105$°. For the RX4 location, the \ac{LOS} component was removed from the measured data at the post-processing stage. The reason for the removal is that the \ac{LOS} component does not interact with the tree for most RX positions as they are not hidden behind the tree, thus it carries no information on the tree. 

\subsection{Measurement and ray-tracing simulation comparison}
\begin{figure}[!t]
\centering
\includegraphics[width=3.2in]{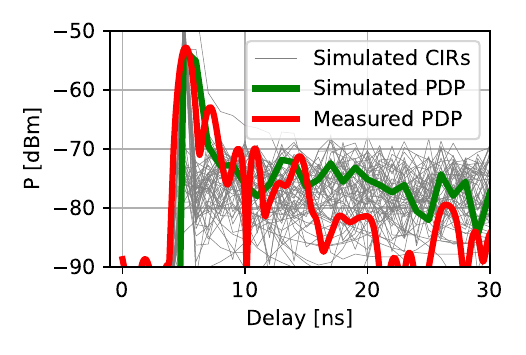}
\caption{{ Measured and ray-traced PDPs and CIRs (in absolute value squared) for $\alpha=0$° at 80\,GHz. We generate twenty random 3D foliage realizations for the given tree crown parameters.}}
\label{fig_PDP}
\end{figure}

In Fig. \ref{fig_PDP} we compare the measured and simulated PDPs at $\alpha=0$° and 80\,GHz. Here we note that the aim of this methodology is not a photorealistic recreation of the tree and subsequent exact \ac{CSI} determination. The aim is a stochastic 3D modeling technique parametrized with physically meaningful parameters, which is used for deterministic and explainable ray-tracing. With that in mind, a qualitative assessment based on visual inspection indicates that the simulation closely matches the measured data.

In Fig. \ref{fig:CDF} we compare \acp{CDF} for both 60 and 80\,GHz. We simulate 20 random realizations of a tree with the density $\rho \in \{0.05, 0.25 \}$, perturbation strength $\xi=0.1$ (i.e. almost circular tree-crown shape), $V_\mathrm{D}=600$\,m$^3$ and $A=2$\,m$^2$. The area of each triangle is notably larger than that of a real-world leaf. However, it was  determined by simulation that reducing the triangle size does not significantly improve the match with measurements. Therefore, the largest triangle area that still provides a good fit to the measurements was used.

In Fig. \ref{fig:CDF} (a) we see the \acp{CDF} of the two frequency bands examined, that are the 60 and 80\,GHz. As for the 60\,GHz band, in Fig. \ref{fig:CDF}\,(a) we read that 50\% of \ac{RSSI} values is -77\,dBm or less. In the case of 80\,GHz in Fig. \ref{fig:CDF}\,(b), we read that 50\% is \ac{RSSI} values are -80\,dBm or less. The difference between the 60 and 80\,GHz is that the 60\,GHz band has approximately 3\,dB lower attenuation, when propagating through a tree. Here it is interesting to remind, that \ac{FSPL} is 2\,dB lower for the 60\,GHz band than for the 80\,GHz. 

\subsection{Angularly-resolved RMS delay spread and path-loss}
A complete description of a wireless channel is provided by the \ac{CIR}, written by \cite{molisch2012wireless}:
\begin{equation}
h(t, \tau) = \sum_{n=1}^N G_n a_n(t,\tau) \delta(\tau - \tau_n),\label{eq:cir}
\end{equation}
where $N$ is the number of \acp{MPC}, $a_n$ is the amplitude and phase of the $n$-th \ac{MPC}. The term $\delta(\tau - \tau_n)$ represents the Dirac delta function, indicating the time delay of the multipath component~$n$.  The effects of the receiving and transmitting antennas are given in the~term~$G_n$.  

\begin{figure*}[!t]
\centering
\subfloat[]{\includegraphics[width=3.2in]{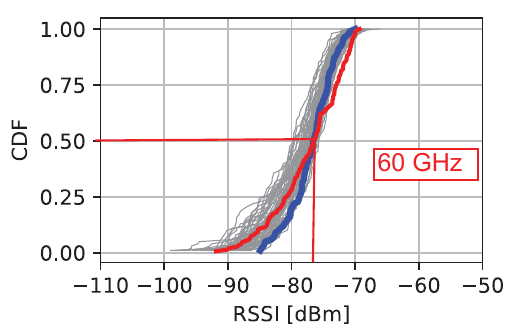}%
\label{fig_first_case}}
\hfil
\subfloat[]{\includegraphics[width=3.2in]{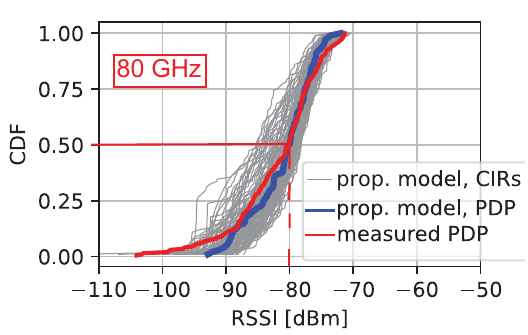}%
\label{fig_third_case}}
\caption{{ CDF evaluation of the simulated and measured PDPs for (a) at 60\,GHz and (b) at 80\,GHz for $\alpha=105$°. Here, the tree crown density was varied such that $\rho \in \{0.05, 0.25 \}$. Please note that the subfigures (a) and (b) share the same legend. }}
\label{fig:CDF}
\end{figure*}

Important parameters of wireless channels are the path-loss and the \ac{RMS} delay spread. The path-loss simply sums all the energy in the \ac{CIR} according to: 
\begin{equation}
\mathrm{PL} = \sum_{\tau} |h(\tau)|^2,
\end{equation}
please note that in the ray-tracing simulation, the \ac{TX} power is set to $0\,\mathrm{dBm}$.

The RMS delay spread describes how strong the delay dispersion of the channel is and it is written as \cite{molisch2012wireless}:  
\begin{equation}{\mathrm{DS}} = \sqrt {\frac{{\sum_\tau {(\tau - {{\bar \tau }})^2P(\tau )} }}{{\sum_\tau {P(\tau )} }}} ,{\text{where }}\bar \tau = \frac{{\sum_\tau {\tau P(\tau )} }}{{\sum_\tau {P(\tau )}}}.
\end{equation}
The term $P(\tau)$ represents the \ac{PDP} and is given~as:
\begin{equation}
P(\tau ) = \mathbb{E}{\left| {h(\tau )} \right|^2}.
\end{equation}
Please note that the simulations and the measurements are assumed to be time-invariant, and thus the Doppler shift is not included in the complex base-band representation of the \ac{CIR} in (\ref{eq:cir}) and for the same reason $P(\tau)$ depends only on $\tau$, while $t$ is omitted.

\begin{figure*}[!t]
\centering
\subfloat[]{\includegraphics[width=3.2in]{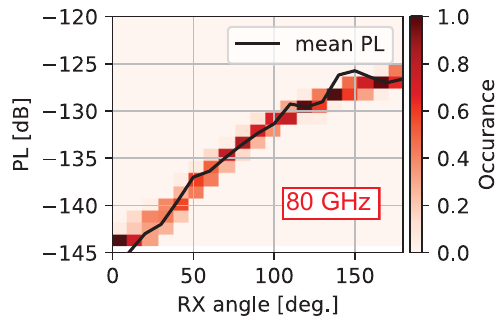}%
\label{fig_pl_60}}
\hfil
\subfloat[]{\includegraphics[width=3.2in]{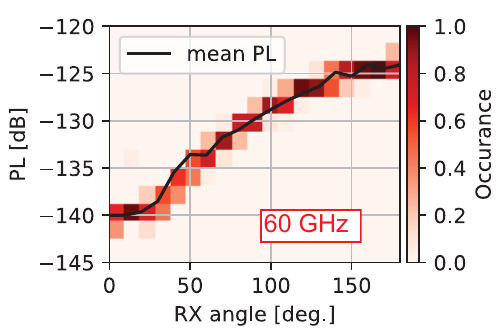}%
\label{fig_pl_80}}
\hfil
\subfloat[]{\includegraphics[width=3.2in]{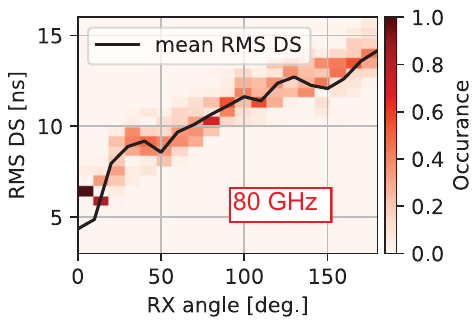}%
\label{fig_ds_80}}
\hfil
\subfloat[]{\includegraphics[width=3.2in]{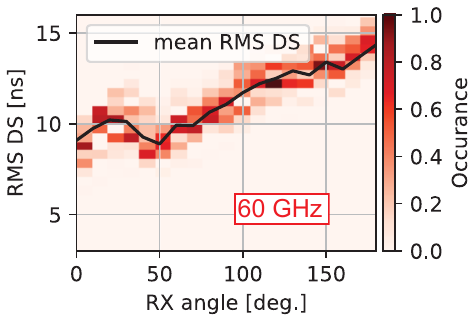}}%
\label{fig_ds_60}
\caption{{ The angularly resolved (a, b) path-loss and (c, d) RMS delay spread heatmaps for the 60 and 80\,GHz bands. For each angle $\alpha$ we generated twenty random 3D realizations of the foliage with the given tree-crown parameters.}}
\label{fig:heatmaps}
\end{figure*}

In Fig. \ref{fig:heatmaps} we can see the evaluation of the ray-tracing simulations for the case when the proposed 3D model of the foliage is orbited by the \ac{RX}. The interpretation of the $\alpha$ angle is available in Fig. \ref{fig:RTscene}. For each $\alpha$, we generate twenty random 3D foliage models using the same settings. Each realization is naturally different, as the tree crown shape, the positions and orientations of the scattering triangles are drawn from random distributions. Thus, in Fig. \ref{fig:heatmaps} we show an occurrence heatmap around the mean value of the examined RMS delay spread and path-loss.

The path-loss is highest for the direct propagation through the tree ($\alpha=0$°). For the 60\,GHz band we read -140\,dB and for the 80\,GHz band the PL is three to four dB higher. Please note that the \ac{LOS} component, if present, would be at -100\,dB at 80\,GHz and -97\,dB at 60\,GHz for the given distance of 30\,m. When we subtract the loss caused by the free space propagation, we see that the additional attenuation caused by the tree is around 20\,dB at $\alpha=180$° and 40\,dB at $\alpha=0$°. In simulations and also in the measurements, the LOS is removed. The reason is that most of the RX locations are not behind the tree, thus the LOS is not interacting with the tree and because it is much stronger than the remaining \acp{MPC}, it would undesirably influence the PL, DS and also the CDFs ~in~Fig.~\ref{fig:CDF}. 

As $\alpha$ increases, the PL almost linearly decreases and for $\alpha=90$° the PL mean value is 133\,dB and 130\,dB for 80\,GHz and 60\,GHz, respectively. This represents an interesting phenomenon, where the strongest \acp{MPC} reflecting back to the transmitter at $\alpha = 180^\circ$ interact only with the first layer of scattering triangles, and are therefore less attenuated than the \acp{MPC} that have to propagate through the entire foliage in the case of $\alpha = 0^\circ$.

The RMS delay spread occurrence heatmap is depicted in Fig. \ref{fig:heatmaps}\,(c) and (d) for 80\,GHz and 60\,GHz, respectively. In general, the DS ranges from 5\,ns to 15\,ns and as $\alpha$ increases, the DS almost linearly increases as well. The intuition behind this is that \acp{MPC} arriving from the direction $\alpha = 180^\circ$ pass through the foliage back and forth when reflecting from the most distant outer layer of scattering triangles. This results in additional delay and an increased delay spread  compared to the direction $\alpha = 0^\circ$. The variance of the DS is approximately 3\,ns and does not change significantly with varying~$\alpha$.

\section{Conclusion}
In this paper we have further developed the 3D foliage modeling methodology and we have shown the angularly-resolved path-loss and RMS delay spread values. We have done this for two frequency bands, i.e. 60\,GHz and 80\,GHz. The ray-tracing simulation, which accepts the proposed 3D model, is done with Sionna, which is an  open-source ray-tracing tool. The simulation outcome is verified by measurements done in a real-world scenario. The PL ranges from -120\,dB to -140\,dB depending on the angle of arrival and angle of departure. The same holds for the RMS delay spread, which ranges from 5\,ns to 15\,ns.

\section*{Acknowledgment}
{The research described in this paper was financed by the Czech Science Foundation, Project No. 23-04304L, "Multi-band prediction of millimeter-wave propagation effects for dynamic and fixed scenarios in rugged time varying environments" and by the National Science Centre, Poland, Project No. 2021/43/I/ST7/03294, through the OPUS-22 (LAP) Call in the Weave Program.}

\bibliographystyle{IEEEtran}
\renewcommand{\IEEEbibitemsep}{0pt plus 0.5pt}
\def\baselinestretch{0.95}
\bibliography{references}

\end{document}